\documentclass[floatfix]{revtex4}
\usepackage{amsmath,amssymb,eucal,mathrsfs}
\usepackage[dvips]{epsfig,graphics}
\usepackage{graphicx}
\usepackage{color}
\usepackage{amsmath}
\usepackage{ulem}

\usepackage[utf8]{inputenc} %% contains utf-8, see: http://inspirehep.net/info/faq/general#utf8

\newcommand{\be}{\begin{equation}}
\newcommand{\ee}{\end{equation}}

\newcommand{\GeV}{\:\text{GeV}}
\newcommand{\TeV}{\:\text{TeV}}

\begin{document}

\title{Testing Higgs boson scenarios in the phenomenological NMSSM}
\author{Shehu~S.~AbdusSalam}
\email{abdussalam@sbu.ac.ir}
\affiliation{Department of Physics, Shahid Beheshti University, Tehran
  19839, Islamic Republic of Iran}

\begin{abstract}
  There could be another scalar in nature quasi-degenerate with the observed one ($h_{125}$).
  This is possible in models such as the Next-to-Minimal Supersymmetric Standard Model (NMSSM).
  The scenario(s) with a 
  single Higgs boson can be compared to that with multiple ones, all near $125 \GeV$. In order to assess
  the extent to which the current set of collider, cold dark matter relic
  density and direct detection limits are capable of discriminating these scenarios, 
  we perform, for the first-time, global fits of
  a weak-scale phenomenological NMSSM with 26 free parameters using the nested
  sampling implementation in {\sc PolyChord}, a next-generation tool for Bayesian inference. The 
  analyses indicate that the data used shows a moderate tendency for supporting the scenario with
    an additional scalar much lighter than $h_{125}$ with mass distribution centred below the W-boson mass. 
  More stringent constraints are, however, needed for decisive inference
  regarding an additional Higgs boson with mass much less than or near $125 \GeV$. 
\end{abstract}

\maketitle

\section{Introduction} 
The Standard Model (SM) of particle physics predicted the existence 
 of a neutral scalar particle, the Higgs boson, with an unknown mass.
 After decades of technological and experimental 
developments since the prediction, eventually such a 
particle with mass near 125 GeV was discovered at
the Large Hadron Collider (LHC) \cite{Aad:2012tfa, Chatrchyan:2012ufa}. This discovery completed
the SM as a successful quantum field theory description of the electroweak and
strong interactions. However, to date, there remained observations and
theoretical indications of physics beyond the SM (BSM). 

Supersymmetry(SUSY)-based BSM, such as the Next-to-Minimal
Supersymmetric Standard model (NMSSM) \cite{Maniatis:2009re,
  Ellwanger:2009dp, Ellwanger:2011sk}, have spectra with multiple 
neutral scalar particles. A lot of theoretical and phenomenological
work  within the NMSSM framework have been done by various groups. To
mention a non-exhaustive selection, these include global
fits of the model's sub-spaces along experimental constraints
\cite{1211.1693,0906.5012,1012.4490,1704.02941, 1703.07800}, studies on its  
relation to baryogenesis \cite{1201.3781, 1706.09898, 1006.1458,
  1102.5679}, phenomenological comparisons \cite{Cao:2012fz, 1509.02452,
  1308.1333, 1207.1096} and 
vacuum stability analyses \cite{1103.5109, 1203.4328, 1612.06803}. In this 
article we analyse, for the first time, a weak-scale NMSSM with many free parameters
in contrast to the analogue at the grand unification theory scale which have been typically studied the most. We call this scenario
the "phenomenological NMSSM" (pNMSSM), as it is constructed with similar
motivations to the well-studied pMSSM~\cite{Djouadi:1998di, Profumo:2004at,
  AbdusSalam:2008uv, Berger:2008cq, AbdusSalam:2009qd}.  

Within the pNMSSM, various Higgs sector scenarios can be considered 
depending on how the observed 125 GeV scalar at the LHC, which we shall label as $h_{125}$, is 
identified and on how the other Higgs boson masses are restricted. Within the literature, 
the scenario studied the most is the ordinary case for which the lightest
CP-even scalar, $h_1$, is identified as $h_{125}$. Here we consider three scenarios,
${\cal H}_0$, ${\cal H}_1$, and ${\cal H}_2$, which are defined as follows:
\begin{itemize}
\item ${\cal H}_0$: $h_1 \equiv h_{125}$. %%  PRL	114	(2015)	191803
  pNMSSM points were discarded if
  $m_{h_1} \notin [122, 128] \GeV$. To make ${\cal H}_0$ mutually exclusive to 
  ${\cal H}_{1,2}$, $m_{h_2} \notin [122, 128] \GeV$ is required for this scenario. 
  No restrictions on the other Higgs bosons were imposed.
\item ${\cal H}_1$: $h_1 \equiv h_{125}$ with $m_{h_2} \in [122, 128] \GeV$.
\item ${\cal H}_2$: $h_2 \equiv h_{125}$. To make ${\cal H}_2$ mutually exclusive to 
  ${\cal H}_{0,1}$, $m_{h_1} \notin [122, 128] \GeV$ is required.
  No restrictions on the other Higgs boson masses were imposed. 
\item ${\cal H}_3$: $h_1 \equiv h_{125}$ with the restriction
  that $m_{h_2} \sim m_{h_3} \sim m_{h_1}$. This and other potentially interesting
  further possibilities will not be addressed in this article.
\end{itemize}
Here $h_i$, with $i = 1,2,3$, represent the three mass-ordered CP-even 
Higgs bosons. 
Our proposal is that by using Bayesian models comparison technique, one can find out 
which of the alternative hypotheses is supported most and, from
another perspective, one can assess the status of the pNMSSM in light of the experimental data used. 
This method has been successfully applied in various particle physics
phenomenology~\cite{0807.4512, AbdusSalam:2009qd, AbdusSalam:2009tr,
  AbdusSalam:2010qp, Bergstrom:2012yi, Bergstrom:2012nx,
  Bergstrom:2014owa, Fowlie:2014xha, Fowlie:2014faa, Fowlie:2016jlx,
  Fowlie:2016rmn, Athron:2017fxj} and to greater extent in cosmology and
astro-particles research. For instance, see \cite{Skilling, astro-ph/0504022, Feroz:2008xx, Handley:2015fda} and
%%  10.1093/mnras/stv1911 = 1506.00171
their citations. For our analyses, we use the {\sc PolyChord}~\cite{Handley:2015fda, mnrasstv1911}
``next-generation'' (to {\sc MultiNest}\cite{Feroz:2008xx}) nested sampling implementation 
for making the comparisons.

The aim of this article is to determine which of the scenarios, ${\cal H}_0$ , ${\cal H}_1$ and ${\cal H}_2$, 
is supported most by current data. Performing such
a comparison analyses will add new directions to previous studies 
concerning quasi-degenerate Higgs boson scenarios such as in 
\cite{Ellwanger:2012ke, Gunion:2012gc, Gunion:2012he, 1509.02452,
  Chen:2016oib, Das:2017tob}. Likewise, the strength of the data in constraining the pNMSSM can be quantified. 
New benchmarks and perhaps experimentally unexplored pNMSSM regions can be extracted 
for future investigations along the searches for BSM physics. In the sections that follow, 
we first describe the Bayesian models comparison technique
which makes the base for our analyses. This will be followed by a description of the
weak-scale parametrisation, the procedure for fitting the parameters to data, and then the results of the comparisons made. 
The article ends with a Conclusions section.

\section{Bayesian model selection} 
Using the nested sampling algorithm, the Bayesian evidence based on a given set of data  
can be readily computed and thereafter used for comparing alternative physics scenarios.
 With this algorithm, the parameters estimation is  
a by-product of the evidence computation. This is a unique advantage of nested sampling in contrast to traditional
Monte Carlo techniques. 

There are various possibilities for performing Bayesian models comparison. 
One can perform a comparison between two completely different physics models based on a common set of data.
For example in \cite{AbdusSalam:2009tr} various SUSY-breaking mediation mechanisms were compared. 
 Another possibility is the comparison of 
alternative physics scenarios within a single model. Examples of this can be
found in \cite{0807.4512, AbdusSalam:2010qp}. In \cite{AbdusSalam:2010qp}, the
comparison was between the MSSM scenario whereby the neutralino
lightest sparticle (LSP) is considered to make all of the observed 
cold dark matter relic density compared to the alternative for which the 
LSP accounts for only part, not all, of the observed relic
density. Along this line of thought, the
pNMSSM scenarios ${\cal H}_0$ , ${\cal H}_1$ and ${\cal H}_2$, can be compared among
one another based on a common set of experimental data. In the subsections
that follow we briefly describe Bayes' theorem, the Bayes factors which were used for the comparisons,
and the similarities/differences between {\sc PolyChord} and {\sc MultiNest}.

\subsection{Bayes theorem and Bayes factors}
For a given context, $\cal{H}$, based on a model with a 
set of $N$ parameters, $ \theta$, the {\it a priori} assumed values which the parameters can take is encoded in the 
prior probability density distribution, $p( \theta )$. The support a given hypothesis could draw from a data 
set is quantified by the probability density of observing the data set given the hypothesis,  
\be \label{evid} 
\mathcal{Z} \equiv p( d|{\cal{H}}) = \int{p( d| \theta,{\cal{H}}) \, p( \theta)}\ d  \theta.
\ee
This can be obtained directly from Bayes' theorem, 
\be \label{bayes}
p( \theta| d, {\cal{H}}) = \frac{p( d|
  \theta,{\cal{H}}) \, p( \theta)}{p( d|{\cal{H}})}.
\ee
Here $p( d| \theta, {\cal{H}})$ is the likelihood, a measure of the probability 
for obtaining the data set $ d$ from a given set of the model 
parameters. To compare between say, ${\cal{H}}_{0}$ and ${\cal{H}}_{1}$  
the Bayes factor, $K = \frac{\mathcal{Z}_1}{\mathcal{Z}_0}$, should be computed. This could be done via the
the posterior odd ratios:
\be \label{postodd}
\frac{p({\cal{H}}_{1}| d)}{p({\cal{H}}_{0}| d)}
=\frac{p( d|{\cal{H}}_{1})p({\cal{H}}_{1})}{p( d| {\cal{H}}_{0})p({\cal{H}}_{0})}
=\frac{\mathcal{Z}_1}{\mathcal{Z}_0}\frac{p({\cal{H}}_{1})}{p({\cal{H}}_{0})}.
\ee
For the case where the two hypotheses are {\it a priori} equally likely,
$p({\cal{H}}_{1})/p({\cal{H}}_{0}) = 1$, then the logarithm of the Bayes factor can be
obtained as the logarithm of the posterior odd factors:
\be \Delta \log_e \mathcal{Z} = \log_e \left[ \frac{p({\cal{H}}_{1}|d)}{p({\cal{H}}_{0}| d)}\right]
= \log_e \left[\frac{\mathcal{Z}_1}{\mathcal{Z}_0}\right].
\ee
Getting $\mathcal{Z}_1/\mathcal{Z}_0 > 1$ will infer that the data supports  
${\cal{H}}_1$ more compared to ${\cal{H}}_0$ and vice versa if the ratio is less than one.
The Jeffrey's scale shown in Table~\ref{tab:Jeffreys} calibrates the significance of the Bayes factors. 
 Next we are going to explain what the Bayes factors describe within the context of the pNMSSM global fits to data. 
 \begin{table}
  \begin{tabular}{|l|l|}
    \hline
    Bayes factors, K &  Comparison Remarks \\ 
    \hline
    1  to 3.2   &  Inconclusive \\
    3.2  to 10  &  Weak Evidence \\
    10 to 100   &  Moderate Evidence \\
    $> 100$     &  Strong Evidence \\
    \hline
  \end{tabular}
  \caption{Jeffreys' scale~\cite{Jeffreys} (see also the work in Ref.~\cite{kassraftery}) for the interpretation of
    the Bayes factors. The Bayes factors, as explained below, could also quantify the relative posterior probability
    masses for the scenarios compared which are a priori equally likely.}
  \label{tab:Jeffreys}
\end{table}

From a point of view, the Bayes factor encodes information about the scenarios' posterior masses 
as measured by the chosen priors. It can tell which scenario is more plausible based on a given set
of data. To see this, let the pNMSSM posterior
without restricting to any of the ${\cal H}_{0}$, ${\cal H}_{1}$, or ${\cal H}_{2}$ scenarios be 
\be p( \theta| d) = \frac{p( d| \theta)\,p( \theta)}{{\cal Z}}. \ee Assuming that the scenarios represent
mutually exclusive volumes, $\Omega_0, \Omega_1$ and $\Omega_2$ respectively, in the ``full'' $\theta$-space then 
the corresponding posterior probability masses can be computed. For instance, 
\be \label{polychordZ}
   {\cal Z}_0^{'} = \int_{\Omega_0} \, p(\theta | d) \, d\theta = \int_{\Omega_0} \, \frac{p( d| \theta)p( \theta)}{{\cal Z}}
    \, d\theta = \frac{{\cal Z}_0}{{\cal Z}}.
\ee
The global evidence, ${\cal Z}$, will cancel out
when computing the ratios, such as $\frac{{\cal Z}_0^{'}}{{\cal Z}_1^{'}} = \frac{{\cal Z}_0}{{\cal Z}_1}$. So the evidence and 
posterior mass ratios are equivalent. As such, 
the priors for scenarios ${\cal H}_{0,1,2}$ are not free to be chosen arbitrarily. They are rather set by the prior
distribution $p(\theta)$, with the relative priors such as $\frac{p({\cal H}_0)}{p({\cal H}_1)}$ 
constrained to match the corresponding 
integrations of $p(\theta|{\cal H}_{0,1})$ over the domains $\Omega_{1,2}$. The prior ratios can be estimated 
by scanning over the pNMSSM parameters without imposing the likelihoods or scenario requirements and then find
the number of survived points after imposing the Higgs boson(s) mass restrictions.
Thus $p({\cal H}_i)$, $i = 0, 1, 2$ can be considered to be the fraction of the survived points. 

For the nested sampling implementation in {\sc PolyChord}, parameter points were sampled from
a flat prior distribution, $p( \theta)$, which integrates to 1 over the ``full'' pNMSSM
$\theta$-space. This ``global'' prior is used for each of the three scenarios considered. So
{\sc PolyChord} is run once for each of the scenarios. From the sampled
parameter points, the ${\cal H}_0$, ${\cal H}_1$, or ${\cal H}_2$ cuts on the Higgs boson masses were applied. 
For instance, in the case of the {\sc PolyChord} run for ${\cal H}_0$ the sampled pNMSSM points were
discarded if $m_{h_1} \notin [122, 128] \GeV$ and ${\cal H}_{1,2}$, $m_{h_2} \notin [122, 128] \GeV$.
As such the evidence value returned by {\sc PolyChord} will be
\be \label{polychordZ}
    {\cal Z}_0^{''} = \int_{\Omega_0} \, p(d | \theta) \, p(\theta) \, d\theta = \int_{\Omega_0} \, p(d | \theta) \,
    p(\theta | {\cal H}_0) \, p({\cal H}_0) \, d\theta
\ee
with $\int \, p(\theta) \, d\theta = 1$, since in principle $p(\theta)$ can be expanded as
$p(\theta) = p(\theta | {\cal H}_0) \, p({\cal H}_0) + p(\theta | {\cal H}_1) \, p({\cal H}_1) + p(\theta | {\cal H}_2) \, p({\cal H}_2)$.
This way, the ratios such as
$\frac{{\cal Z}_0^{''}}{{\cal Z}_1^{''}} = \frac{{\cal Z}_0}{{\cal Z}_1} \,\, \frac{p({\cal H}_0)}{p({\cal H}_1)}$
represents the full posterior mass ratio as in Eq.(\ref{postodd}). Thus the Bayes factor, K can be obtained from
what {\sc PolyChord} returns (the ${\cal Z}^{''}$s) as
\be K = \frac{{\cal Z}_0}{{\cal Z}_1} = \frac{{\cal Z}_0^{''}}{{\cal Z}_1^{''}}   \,\,  \frac{p({\cal H}_1)}{p({\cal H}_0)}. \ee

\subsection{{\sc PolyChord} versus {\sc MultiNest}}
Here we briefly describe the similarities and contrast 
 between the relatively new {\sc PolyChord} \cite{Handley:2015fda, mnrasstv1911}
 and the {\sc MultiNest} algorithm \cite{Feroz:2007kg,
  Feroz:2008xx} which we have used in the past for similar analyses. Both {\sc MultiNest} and
{\sc PolyChord} are effective Bayesian evidence calculators that perform as 
excellent multi-modal posterior samplers. At their core, ``nested sampling''
algorithm \cite{Skilling} is implemented. They differ on how a new set of model parameters
is generated over sampling iterations. {\sc MultiNest} is based on 
rejection sampling or, alternatively, importance sampling. On the other hand, {\sc PolyChord} is based
on slice sampling method. 

We used {\sc PolyChord} because of its improved scaling with dimensionality, $D$,
as illustrated in \cite{mnrasstv1911}. For the multi-dimensional Gaussian problem
analysed in \cite{mnrasstv1911}, the number of likelihood calculations needed for the run to
converge scales as ${\cal O}(D^3)$ at worst using {\sc PolyChord} instead of
the approximately exponential scaling that emerges for higher dimensions
(greater than around 20) as is expected
for the rejection sampling method (see section 29.3 of \cite{Mackay}). 
 The 26-dimensional pNMSSM considered for the analysis presented
in this article is more complicated in comparison to the toy Gaussian problem.

There are two tuning parameters for running {\sc PolyChord}, namely the
number of live points maintained throughout the nested sampling implementation,
$n_{live}$, and the length of the slice sampling chain, $n_{repeats}$, used for
generating new live points. With $n_{live} = 200$ and $n_{repeats} = 26$, 
running {\sc PolyChord} on the pNMSSM parameters space finished with 700208
likelihood calculations using 96 core-hours of computing time.
This compares to 11344428 likelihood calculations for the same pNMSSM model
with {\sc MultiNest} tuning parameters $n_{live} = 5000$ and $efr = 0.1$ using
4480 core-hours. Making $n_{live} = 1000$ instead of $n_{live} = 200$, the amount
of time and likelihood calculations needed to finish the {\sc PolyChord} run
increased drastically but ends up with similar results for the Bayesian evidence.
It took 6272 core-hours and 3492331 likelihood calculations.
Setting $n_{repeats} = 2 \times 26$ instead of $n_{repeats} = 26$, {\sc PolyChord}
finished with 1632936 likelihood calculations and 512 core-hours. These basic
comparison between {\sc PolyChord} and {\sc MultiNest} are summarised in
Table~\ref{tab:comp}. Given the experience in using {\sc MultiNest} for
large parameter models (order 20 to 30), especially the difficulties in
getting runs to finish over tightly constrained or models with high number of parameters,
 we decide to use {\sc PolyChord}. \footnote{Note: for the comparison performed here, the aim
  is about getting correct evidence values. No special attention were made for comparing the
  uncertainties on the Evidence coming from the codes.
  The {\sc PolyChord}'s ``precision criterion'' used is $10^{-3}$ while the {\sc MultiNest}'s 
  ``tolerance'' parameter was set to 0.5. These code parameters affect, but in different ways
  (see section 5.4 of \cite{Feroz:2008xx}),
  the stopping criterion for the nested sampling and the error on the evidence. For the basic
  comparison made here between the codes no attempt were made for achieving similar
  deviations on the evidence values.}
\begin{table}
  \begin{tabular}{|l|l|l|l|l|l|}
    \hline
        {\sc MultiNest} & $N$ & $efr$ & $\log \mathcal{Z}$ & $t_{CPU}$ [core-hours] & $N_l$ \\
    \hline
    & 5000 & 0.8 & $5.77 \pm 0.05$ & 64 & 84195 \\
    & 5000 & 0.1 & $6.30 \pm 0.04$ & 4480 & 11344428 \\
    \hline
        {\sc PolyChord} & $N$ & $r$ & $\log \mathcal{Z}$ & $t_{CPU}$ [core-hours] & $N_l$ \\      
    \hline
    & 200 & 26 & $6.20 \pm 0.22$ & 96 & 700208 \\
    & 1000 & 26 & $5.83 \pm 0.10$ & 6272 & 3492331 \\
    & 200 & $2 \times 26$ & $6.11 \pm 0.22$ & 512 & 1632936 \\
    \hline
  \end{tabular}
  \caption{An example of basic quantitative comparison of the relative performance
    between {\sc PolyChord} and {\sc MultiNest} for the 26 parameters pNMSSM.
    Here $N$, $efr$, $r$, $\log \mathcal{Z}$, $t_{CPU}$ and $N_l$ are  
    respectively the nested sampling number live points, {\sc MultiNest} algorithm
    tuning parameter, {\sc PolyChord} tuning parameter ($n_{repeats}$), the logarithm of
    the Bayesian factor obtained in a run, the CPU core-hours taken and the number of 
    likelihood calculations done before finishing a run.}
  \label{tab:comp}
\end{table}

\section{The phenomenological NMSSM} 
The NMSSM, for reviews see e.g. \cite{Maniatis:2009re, Ellwanger:2009dp, Ellwanger:2011sk}, 
has phenomenological advantages over the MSSM. These  
include the solution of the $\mu$-problem \cite{Kim:1983dt}. The vacuum expectation
value of an additional gauge-singlet (S) can generate superpotential
$\mu$-term dynamically. It also has a richer Higgs-sector. There are 
three CP-even Higgs bosons, $h_{1,2,3}$, and two CP-odd Higgs bosons $a_{1,2}$
which are mixtures of the MSSM-like Higgs doublet fields and respectively the real or
imaginary part of $S$. For our analyses, we shall consider an R-parity conserving NMSSM with
minimal CP and flavour violating free parameters and superpotential, 
\be \label{superpot}
W_{NMSSM} = W_{MSSM'} - \epsilon_{ab}\lambda {S} {H}^a_1 {H}^b_2 + \frac{1}{3}
\kappa {S}^3 \ ,
\ee 
where $W_{MSSM'}$ is the MSSM-like superpotential without the $\mu$-term, 
\begin{eqnarray}
  W_{MSSM'}&=& \epsilon_{ab} \left[ 
    (Y_E)_{ij} H_1^a    L_i^b    {\bar E}_j 
    + (Y_D)_{ij} H_1^a    Q_i^{b}  {\bar D}_{j} 
    + (Y_U)_{ij} H_2^b    Q_i^{a}  {\bar U}_{j}
    \right].
  \label{wmssm}
\end{eqnarray}
Here, the chiral superfields have the following $SU(3)_C\otimes SU(2)_L\otimes U(1)_Y$ quantum numbers, 
\begin{eqnarray} 
  L:(1,2,-\frac{1}{2}),\quad {\bar E}:&(1,1,1),\quad\, Q:\,(3,2,\frac16),\quad
  {\bar U}:\,(\bar{3},1,-\frac{2}{3}), {\bar D}:&(\bar{3},1,\frac13),\quad
  H_1:(1,2,-\frac{1}{2}),\quad  H_2:\,(1,2,\frac{1}{2}). 
  \label{fields}
\end{eqnarray}
The corresponding soft SUSY-breaking terms are
\be
V_\mathrm{soft} = V_2 + V_3 + m_\mathrm{S}^2 | S |^2 +
(-\epsilon_{ab}\lambda A_\lambda {S} {H}^a_1 {H}^b_2 + 
\frac{1}{3} \kappa A_\kappa {S}^3  
+ \mathrm{H.c.}), \ee
with the trilinear and bilinear contributions given by 
\begin{eqnarray}
  V_2 & = & m_{H_1}^2 {{H^*_1}_a} {H_1^a} + m_{H_2}^2 {{H^*_2}_a}
  {H_2^a} + 
  {\tilde{Q}^*}_{i_La} (m_{\tilde Q}^2)_{ij} \tilde{Q}_{j_L}^{a} +
  {\tilde{L}^*}_{i_La} (m_{\tilde L}^2)_{ij} \tilde{L}_{j_L}^{a}  
  + \nonumber \\ &&
  \tilde{u}_{i_R} (m_{\tilde u}^2)_{ij} {\tilde{u}^*}_{j_R} +
  \tilde{d}_{i_R} (m_{\tilde d}^2)_{ij} {\tilde{d}^*}_{j_R} +
  \tilde{e}_{i_R} (m_{\tilde e}^2)_{ij} {\tilde{e}^*}_{j_R}, \\
  V_3 & = & \epsilon_{ab} \sum_{ij}
  \left[
    (T_E)_{ij} H_1^a \tilde{L}_{i_L}^{b} \tilde{e}_{j_R}^* +
    (T_D)_{ij} H_1^a \tilde{Q}_{i_L}^{b}  \tilde{d}_{j_R}^* +
    (T_U)_{ij}  H_2^b \tilde{Q}_{i_L}^{a} \tilde{u}_{j_R}^*
    \right]
  + \mathrm{H.c.}. 
\end{eqnarray}
A tilde-sign over the superfield symbol represents the scalar
component. However, an asterisk over the superfields as in, for example, 
$\tilde{u}_R^*$ represents the scalar component of $\bar{U}$. 
The $SU(2)_L$ fundamental representation indices are donated by
$a,b=1,2$ while the generation indices by
$i,j=1,2,3$. $\epsilon_{12}=\epsilon^{12}=1$ is a totally
antisymmetric tensor. 
In a similar approach to the pMSSM~\cite{Djouadi:1998di, AbdusSalam:2008uv,
  Berger:2008cq, AbdusSalam:2009qd} construction, the pNMSSM parameters are
defined at the weak scale. For
suppressing sources of unobserved CP-violation and flavour-changing
neutral currents, the sfermion mass and trilinear scalar coupling
parameters were chosen to be real and diagonal. For the same motivation, the
first and second generation sfermion mass parameters were set to be
degenerate. The gaugino mass
parameters were reduced to be real by neglecting
 CP-violating phases. These lead to a non-Higgs sector set of parameters
\be \label{susypart} M_{1,2,3};\;\; m^{3rd \,
  gen}_{\tilde{f}_{Q,U,D,L,E}},\;\;  m^{1st/2nd \,
  gen}_{\tilde{f}_{Q,U,D,L,E}}; \;\;A_{t,b,\tau}. 
\ee 
Here, $M_{1,2,3}$ and $m_{\tilde f}$ are respectively the gaugino and
the sfermion mass parameters. $A_{t,b,\tau}$ represent the
trilinear scalar couplings, with $T_{ij} \equiv A_{ij} Y_{ij}$ (no
summation over $i,j$). So $A_{t,b,\tau}$ is equivalent to the $A_{33}$
corresponding, respectively, to the diagonalised matrices $T_{U}$, $T_{D}$, and
$T_{L}$. Here $Y$ represent the Yukawa matrices.
After electroweak symmetry breaking, the
vacuum expectation value (vev) of $S$, $v_s$, develops an effective  $\mu$-term,   
$\mu_\mathrm{eff} = \lambda \, v_s$. This and the ratio of the MSSM-like Higgs doublet 
vevs, $\tan \beta=\left<H_2\right>/\left<H_1\right>$, are  
free parameters which together with mass of the Z-boson, $m_Z$, can be
used for computing $m^2_{H_{1,2},S}$ via minimisation of the scalar
potential. 
With these, the tree-level Higgs sector parameters are
  \be \label{higgspars}
\tan\beta, \lambda, \kappa, A_{\lambda}, A_{\kappa}, \lambda \, v_s.  
\ee
Adding to the list of parameters in Eq.(\ref{susypart}) and Eq.(\ref{higgspars}), 
four SM nuisance parameters, namely, the top and bottom quarks $m_{t,b}$, $m_Z$ 
and the strong coupling constant, $\alpha_s$, 
makes the 26 free parameters of the pNMSSM:
\be \label{theparams}
 \theta = \{ M_{1,2,3};\;\; m^{3rd \, gen}_{\tilde{f}_{Q,U,D,L,E}},\;\; 
m^{1st/2nd \, gen}_{\tilde{f}_{Q,U,D,L,E}}; \;\;A_{t,b,\tau,\lambda, \kappa};
\;\; \tan \beta, \lambda, \kappa, \mu_\mathrm{eff}; \;\; m_{t,Z,b},
\alpha_s \}.   
\ee

$M_{1,2}$ strongly affect the
electroweak gaugino masses for which a wide range of values, GeV to TeV, is 
possible. We let $M_1 \in [-4, 4] \TeV$ and same for $M_2$ but
fixed to be positive without loss of generality (see e.g. \cite{Choi:2001ww}). 
A strong sensitivity of the pNMSSM Higgs sector on the gluino and the
1st/2nd generation squark mass parameters is not anticipated. However we choose
to let them vary since the limits from searches for SUSY will be part of the experimental
data to be used. As such, following the work in \cite{AbdusSalam:2008uv,
  AbdusSalam:2009qd} we let the gluino and squark mass parameters vary
within $[100 \GeV, 4 \TeV]$ and the trilinear scalar couplings 
 within $[-8 \TeV, 8 \TeV]$. $\tan \beta$ is
allowed to vary between 2 and 60. With the aim of minimising
fine-tuning, we subjectively choose to vary the effective
$\mu$-parameter, $\mu_\mathrm{eff} = \lambda \, v_s$, to be within 100
to 400 GeV and not (orders of magnitude) far away from the $m_Z$. The
remaining Higgs-sector parameters were allowed in ranges as summarised
in Table~\ref{tab.param}. For all the parameters, except the SM ones, flat
prior probability density distribution was assumed. 
For the experimentally measured SM nuisance parameters, Gaussian distributions around
the measured values were used. 
\begin{table}[htbp!] 
  \begin{center}{\begin{tabular}{|ll|}
        \hline
        Parameter & Range\\
        \hline
        $M_1$                   & [$-4$ TeV, $4$ TeV]\\
        $M_2$                   & [$0$ TeV, $4$ TeV]\\
        $M_3, \,\, m^{3rd \, gen, \,\, 1st/2nd \, gen}_{\tilde{f}_{Q,U,D,L,E}}$   & [$100$ GeV, $4$ TeV]\\   
        $A_{t,b,\tau}$ &  [-$8$ TeV, $8$ TeV]\\ 
        %%        $\tan \beta$ &  [$2$, $15$] \\
        $\tan \beta$ &  [$2$, $60$] \\ 
        $\lambda$ & [$10^{-4}$, $0.75$] \\
        $\kappa$ & [$-0.75$, $0.75$] \\
        $\mu_\mathrm{eff}$ & [$100$, $400$] GeV\\
        $A_{\lambda}$ &  [$50$ GeV, $4$ TeV]\\
        $A_{\kappa}$ &  [$-2$ TeV, $2$ TeV]\\ 
        \hline
        $m_t$  &  172.6 $\pm$ 1.4 GeV\\
        $m_Z$  &  91.1876 $\pm$ 0.0021 GeV\\
        $m_b(m_b)^{\overline{MS}}$  &  4.20 $\pm$ 0.07 GeV\\
        $\alpha_{s}(m_Z)^{\overline{MS}}$ & 0.1172 $\pm$ 0.002 \\ 
        \hline
  \end{tabular}}\end{center}
  \caption{The 26 pNMSSM parameters and their corresponding flat prior
    probability density distribution ranges. The SM parameters were varied according to
    Gaussian distributions with the shown central values and standard deviations. 
  \label{tab.param}}
\end{table}

Checking the prior-dependence of results is useful for assessing the strength of the data
in constraining the model in an unambiguous manner. 
The priors can be chosen to be flat or
logarithmic with the latter favouring lower regions of the parameter ranges. There are
two bottle-necks concerning our attempts for sampling the pNMSSM parameters with 
logarithmic priors. On one hand, the absence of signatures for SUSY at the LHC
pushes sparticle mass lower bounds towards or well into the multi-TeV regions. Therefore, 
sampling the phenomenologically viable parameters according to a logarithmic prior 
 will be difficult and computationally expensive. For the attempted log-prior fits,
only parameters that do not cross zero were sampled logarithmically. Those that have the
possibility of being zero were sampled uniformly.
On the other hand, for the nested sampling 
algorithm in {\sc PolyChord} to get started, a 200-points sample of the
26-parameters pNMSSM is required. By using logarithmic priors, it was not
possible to generate the 200 model points within the maximal possibility of 3072 CPU core-hours per
run at our disposal~\footnote{This limitation is because {\sc PolyChord} requires live points to be
  generated and then the start of nested sampling all over a single run of the program.}. Thus we restrict 
our analyses to the flat priors only. The conclusions presented in this article are valid only
within this context. 

Another issue concerning our pNMSSM parametrisation is related to 
``naturalness'' or the avoidance of excessive fine-tuning associated with
obtaining the correct weak-scale ($m_Z$). 
With the naturalness prior parametrisation~\cite{Allanach:2006jc}, the fine-tuning can be avoided
by directly scanning the parameters $m_Z$ and $\tan \beta$ rather than $m_{H_1}$ an $m_{H_2}$. 
Moreover it was shown \cite{Athron:2017fxj} that naturalness prior
parametrisation could significantly affect the Bayesian evidence values. For the analyses
presented in this article, such parametrisation was not considered. Rather, a
Gaussian prior for $m_Z$ centred on the measured value was used. Doing this injects
information about what the weak scale is into the prior. In addition, 
 $\mu_{eff}$ were chosen to be near
the electroweak symmetry breaking scale, between 100 to 400 GeV, 
since one of our aims is to show that there are still vast regions in parameter
space with low mass gauginos and BSM Higgs bosons which are not ruled
out by current data. It is accepted that fine-tuning penalisation manifests implicitly 
and automatically within Bayesian global fits as presented in \cite{Cabrera:2008tj,
  Ghilencea:2012qk, Fowlie:2014xha}. The same applies to the fact that fine-tuning limits
could be imposed during fits by using the various fine-tuning measures 
\cite{Ellis:1986yg, Barbieri:1987fn, Harnik:2003rs,
  Kitano:2005wc, Allanach:2006jc, Baer:2013gva, AbdusSalam:2015uba}. Using
any of these is not within the scope of our present analyses given
the deliberate target to regions with low mass electroweak gauginos. 

Now, coming back to the Higgs sector potential, the details concerning the NMSSM Higgs
mass matrices and couplings can be found in
the literature, for example see \cite{Ellis:1988er, Drees:1988fc, Franke:1995tc, Miller:2003ay,
  Ellwanger:2004xm, Ellwanger:2009dp}. 
The $\mathbb{Z}_3$-invariant NMSSM Higgs potential can be obtained from the SUSY gauge
interactions, soft-breaking and $F$- terms as 
\begin{eqnarray}
V_\mathrm{Higgs} &=& \left|\lambda H_2.H_1 + \kappa S^2 \right|^2 + \left(m_{H_2}^2 + \left|\lambda S\right|^2\right) 
\left(\left|H_2^0\right|^2 + \left|H_2^+\right|^2\right) 
+\left(m_{H_1}^2 + \left|\lambda S\right|^2\right) 
\left(\left|H_1^0\right|^2 + \left|H_1^-\right|^2\right) \nonumber \\
&&+\frac{g_1^2+g_2^2}{8}\left(\left|H_2^0\right|^2 +
\left|H_2^+\right|^2 - \left|H_1^0\right|^2 -
\left|H_1^-\right|^2\right)^2
+\frac{g_2^2}{2}\left|H_2^+ H_1^{0*} + H_2^0 H_1^{-*}\right|^2 +m_{S}^2 |S|^2
+\big( \lambda A_\lambda H_2.H_1 S \nonumber \\
&& +  \frac{1}{3} \kappa A_\kappa\, S^3 + \mathrm{H.c.}\big), \textrm{ with }
H_1 = \left( \begin{array}{c} H_1^0 \\ H_1^- \end{array} \right), \textrm{ and }
H_2 = \left( \begin{array}{c} H_2^+ \\ H_2^0 \end{array} \right).
\end{eqnarray}
Here $g_1$ and $g_2$ denotes the $U(1)_Y$ and $SU(2)$ gauge couplings,
respectively. Out of the 22 non-SM pNMSSM parameters, six, compared to two for the pMSSM, are directly from
the Higgs sector. After electroweak symmetry breaking, replacement of the Higgs
sector fields with corresponding fluctuations on top of the vevs,
\begin{equation}
H_2^0 = \left<H_2\right> + \frac{H_{2R} + iH_{2I}}{\sqrt{2}}, \quad
H_1^0 = \left<H_1\right> + \frac{H_{1R} + iH_{1I}}{\sqrt{2}}, \quad
S = s + \frac{S_R + iS_I}{\sqrt{2}}\;,
\end{equation}
leads to the realisation of CP-even Higgs boson mixing matrix 
\begin{equation}
h_i^{\text{mass}} = O_{ij} h_j^{\text{weak}}. 
\end{equation}
Here the physical Higgs fields have indices~R for the CP-even, and indices~I for
the CP-odd states. $h_i^{\text{weak}} = (H_{1R}, H_{2R}, S_R)$ represents the interaction, and
$h_i^{\text{mass}}$, the mass-ordered, eigenstates. The mixing of the $SU(2)$
doublets with the singlet state affects the phenomenology of the Higgs bosons. For
instance, the reduced couplings (see, e.g. \cite{Ellwanger:2004xm})
\begin{equation} 
\xi_i = \sin\beta\, O_{i2} + \cos\beta\, O_{i1} 
\end{equation} 
of the 3 CP-even mass-eigenstates $h_i$ to the electroweak gauge bosons can be very
small in some regions of parameter space. The sum rule $\sum_{i=1}^3 \xi_i^2 = 1$ 
is always satisfied. The reduced couplings are inputs to the {\sc Lilith}~\cite{Bernon:2015hsa}
program for comparing the pNMSSM signal strengths to the experimentally measured values.

\section{Experimental constraints and fit procedure}
During the global fits, the experimental constraints used were those implemented in {\sc NMSSMTools} \cite{Ellwanger:2006rn, 
  Ellwanger:2005dv, Djouadi:1997yw, Degrassi:2009yq, Domingo:2007dx, Domingo:2015wyn},
{\sc Lilith} \cite{Bernon:2015hsa}, and {\sc  MicrOMEGAs} \cite{Boos:2004kh, 
  Semenov:2008jy, Belanger:2010st, Pukhov:1999gg, Belyaev:2012qa,
  Belanger:2013oya, Belanger:2010pz, Belanger:2008sj, Belanger:2006is, 
  Barducci:2016pcb}. The set of experimental constraints, $ d$, shown in
Table~\ref{tab.obs} were used to associate each pNMSSM point, $\{ \theta, {\cal{H}}\}$, 
with a likelihood $p( d| \theta, {\cal{H}})$. The likelihood as a function of the 
parameters is explained as follows. 
\begin{table}
  \begin{center}{\begin{tabular}{|lll|}
        \hline
        Observable & Constraint   & References\\ 
        \hline
        $m_h$& $125.09 \pm 3.0$ GeV & \cite{Aad:2015zhl} \\
        $Br(B \rightarrow X_s \gamma)$ & $(3.32 \pm 0.16) \times
        10^{4}$ & \cite{Bobeth:1999ww, Buras:2002tp, Amhis:2014hma}\\  
        $Br(B_s \rightarrow \mu^+ \mu^-)$ & $(3.0 \pm 0.6) \times
        10^{-9}$ & \cite{Aaij:2017vad, Bobeth:2013uxa, Buras:2002vd}  \\ 
        $\Delta M_{B_s}$ & $17.757 \pm 0.021$ & \cite{Buras:2002vd,Ball:2006xx}\\
        $\Delta M_{B_d}$ & $0.5064 \pm 0.0019$ &   \cite{Buras:2002vd,Ball:2006xx}\\ 
        $Br(B_u \rightarrow \tau \nu)$ & $1.06 \pm 0.19$ &
        \cite{Barate:2000rc, Aubert:2004kz, Gray:2005ad, Akeroyd:2003zr}\\
        $\delta a_{\mu}$ & $(30.2 \pm 8.80) \times 10^{-10}$ &
        \cite{Bennett:2006fi, Domingo:2007dx, Domingo:2015wyn} \\        
        $\Omega_{CDM} h^2$ & $0.12 \pm 0.02$ & \cite{Ade:2015xua} \\
        Higgs signal strengths  & & %% in {\sc Lilith} v1.1 with data v15.09 & & 
        \cite{Aaltonen:2013xpo, Aad:2015gba, Aad:2014eha,
          Aad:2015ona, Aad:2014eva, Aad:2015vsa, Aad:2015iha, Aad:2015gra,
          Aad:2014xzb, Aad:2014xva, ATLAS-CONF-2015-004, Aad:2014iia,
          Khachatryan:2014jba, Khachatryan:2014ira, Chatrchyan:2013iaa,
          Chatrchyan:2013mxa, Chatrchyan:2014nva, Chatrchyan:2013zna,
          Khachatryan:2014qaa, Khachatryan:2015ila, Khachatryan:2015bnx,
          Chatrchyan:2014tja} \\
        CDM direct detection limits & &  %% (DD) limits in {\sc NMSSMTools}  & & 
        \cite{Akerib:2016vxi, Aprile:2017iyp, Tan:2016zwf, Amole:2015pla, Amole:2016pye, Akerib:2016lao, Fu:2016ega}\\
        \hline
        \hline
        Constraints in {\sc HiggsBounds} & & \cite{Bechtle:2011sb, Bechtle:2013wla,
          arXiv:1402.3051, arXiv:1509.04670, arXiv:1402.3244, arXiv:1404.1344, arXiv:1406.7663,
          arXiv:1504.00936, arXiv:1409.6064, arXiv:1507.05930, arXiv:1406.5053,
          Khachatryan:2014jba, Khachatryan:2014ira, arXiv:1509.00389}\\
        Constraints in {\sc SModelS} & & \cite{Ambrogi:2017neo,Kraml:2013mwa,
          Buckley:2013jua, Sjostrand:2006za, Beenakker:1996ch, Beenakker:1997ut,
          Kulesza:2008jb, Kulesza:2009kq, Beenakker:2009ha, Beenakker:2010nq, Beenakker:2011fu, 
          ATLAS-CONF-2013-024, ATLAS-CONF-2013-047, ATLAS-CONF-2013-053, ATLAS-CONF-2013-061, CMS-PAS-SUS-13-018, CMS-PAS-SUS-13-023,
          Chatrchyan:2013lya, Khachatryan:2014qwa, Chatrchyan:2014lfa, Khachatryan:2015vra, Chatrchyan:2013xna}\\
        \hline 
  \end{tabular}}\end{center}
  \caption{Summary of the data used for the pNMSSM global fits. Theoretical uncertainties have been added in quadrature
    to the experimental uncertainties quoted. The Higgs signal strengths are those implemented in {\sc Lilith} which
    returns a log-likelihood value for each model point. The cold dark matter direct detection limits implemented are shown in
    Fig.~\ref{dmdetection}. Finally, the {\sc HiggsBounds} and {\sc SmodelS} constraints impose the 95\% C.L. bounds
    on collider pseudo-observables such as $\sigma.Br$ for specific Higgs and SUSY processes respectively.}
  \label{tab.obs}
\end{table}
In modelling the likelihood, the set of constraints used during the global fits are divided into the three groups: 
\begin{itemize}
\item Constraints on the Higgs boson mass $m_h$, the neutralino cold
  dark matter (CDM) relic 
  density $\Omega_{CDM} h^2$, anomalous magnetic moment of the muon 
  $\delta a_\mu$ and B-physics related limits summarised in the upper part of Table~\ref{tab.obs} form
  the first part of the data set, $ d$. The likelihood is computed from the pNMSSM predictions,
  $O_i$, corresponding to the constraints $i$, with experimental central values $\mu_i$ and uncertainties $\sigma_i$, as 
  \be \label{likel} p( d| \theta, {\cal{H}}) = \prod_i \,
  \frac{ \exp\left[- (O_i - \mu_i)^2/2 \sigma_i^2\right]}{\sqrt{2\pi
      \sigma_i^2}}.
  \ee   
  Here the index $i$ runs over the relevant experimental constraints in Table~\ref{tab.obs}. 
\item Signal strength measurements from Tevatron
  \cite{Aaltonen:2013xpo}, ATLAS \cite{Aad:2015gba, Aad:2014eha,
    Aad:2015ona, Aad:2014eva, Aad:2015vsa, Aad:2015iha, Aad:2015gra,
    Aad:2014xzb, Aad:2014xva, ATLAS-CONF-2015-004, Aad:2014iia} and
  CMS \cite{Khachatryan:2014jba, Khachatryan:2014ira,
    Chatrchyan:2013iaa, Chatrchyan:2013mxa, Chatrchyan:2014nva,
    Chatrchyan:2013zna, Khachatryan:2014qaa, Khachatryan:2015ila,
    Khachatryan:2015bnx, Chatrchyan:2014tja} as
  implemented in {\sc Lilith} v1.1 (with data version 15.09)
  \cite{Bernon:2015hsa} represent the second part of the data set. For each pNMSSM point, the
  returned likelihood from {\sc Lilith} is combined with the product in Eq.(\ref{likel}).
The likelihood can be computed via either the Higgs boson signal strengths
  or their reduced couplings with respect to the SM. For the first case, a pNMSSM point
  with corresponding signal strength $\mu_i$ is associated with the likelihood
\be
-2L_{lilith}( \theta) = - 2 \sum_i \log L(\mu_i) = \sum_{i}
\left(\frac{\mu_i(\theta) - \hat\mu_i}{\Delta \mu_i}\right)^2.  
\ee
Here $i$ runs over the various categories of Higgs boson production and decay modes 
combinations. $\hat\mu_i \pm \Delta\hat\mu_i$ represents the experimentally determined signal 
strengths. Theoretically, the signal strength associated to a model point can be computed, for a given production mode $X$ and decay mode $Y$ as
\be \label{muimp}
\mu = \sum_{X,Y} \epsilon_{X,Y} \frac{ \sigma(X) \, BR(H \rightarrow Y)}{\left[ \sigma(X) \, BR(H \rightarrow Y) \right]^{SM}}. 
\ee
Here $\epsilon_{X,Y}$ are the experimental efficiencies, $X \in \{ ggH, VH, VBF, ttH\}$ and
$Y \in \{ \gamma \gamma, VV^{(*)}, b \bar{b}, \tau \tau, ttH\}$.
For a proton-proton collider, the elements in $X$ represent: the gluon-gluon 
fusion (ggH), associated production with a boson (VH), vector boson fusion (VBF)
or associated production with top quarks (ttH). The elements in $Y$ represent the Higgs diphoton 
($\gamma \gamma$), W or Z bosons ($VV$), bottom quarks ($bb$) or tau leptons ($\tau \tau$) 
decay modes.

Now, computing $\mu$ as in Eq.(\ref{muimp}) could be impractical since for a 
meaningful theory versus experiment comparison, the non-SM predictions in the numerator 
should be computed using the same prescriptions such as the order in perturbation,
implementation of 
parton distribution functions etc. The second approach, whereby the input to {\sc Lilith}
are the reduced couplings does not suffer from this problem. BSM physics
effects can be parametrised in terms of the reduced couplings. The cross section (or partial 
decay width) for each production process $X$ (or decay mode $Y$) can be scaled \cite{Heinemeyer:2013tqa} 
with a factor of $C_X^2$ and $C_Y^2$ respectively such that 
\be 
\sigma(X) =  C_X^2 \, \sigma(X)^{SM} \quad \textrm{ and } \quad \Gamma(Y) =  C_Y^2 \, \Gamma(Y)^{SM}. 
\ee 
The reduced couplings computed from the {\sc NMSSMTools} together with
their invisible and undetectable decay branching ratios can then be passed to {\sc Lilith} for 
computing the likelihood based on
\be 
\mu = ( 1 - BR(H \rightarrow  \, undetected) - BR(H \rightarrow \, invisible) ) \frac{ \sum_{X,Y} \epsilon_{X,Y} C_X^2 \, C_Y^2}
    {\sum_Y \, C_Y^2 \, BR(H \rightarrow Y)^{SM}}
\ee
and the table of likelihood values as a function of $\mu$ within the {\sc Lilith} database of 
experimental results. This procedure is valid only for Higgs bosons with mass between 123 to 128 GeV. 
For the multi-Higgs case with masses within this range, such as for ${\cal H}_1$, 
the combined \cite{Bechtle:2015pma} signal strengths %% can be summed incoherently \cite{Bechtle:2015pma} and then the resultant 
were used. 

\item The third set of constraints in $d$ is the CDM direct detection limits. These are from searches for the elastic scattering of
  CDM with nucleons. The recoil energy deposited on nuclei in a detector can be measured.
  In the absence of discovery, upper limits on the scattering cross section can be determined.
  The cross sections can be either spin-independent (SI)
  or spin-dependent (SD) depending on whether the LSP-nucleon coupling is via scalar or axial-vector interaction.
  For the fits with the direct detection limits imposed, only parameter
  points that pass the SI~\cite{Akerib:2016vxi, Aprile:2017iyp, Tan:2016zwf}
  and SD~\cite{Amole:2015pla, Amole:2016pye, Akerib:2016lao, Fu:2016ega} limits were accepted.    
\end{itemize}

Another set of limits were used for fitting the pNMSSM. These were not 
included during the global fit samplings.  Instead, the limits implemented in 
{\sc SModelS}~\cite{Ambrogi:2017neo,Kraml:2013mwa,
  Buckley:2013jua, Sjostrand:2006za, Beenakker:1996ch,
  Beenakker:1997ut, Kulesza:2008jb, Kulesza:2009kq, Beenakker:2009ha,
  Beenakker:2010nq, Beenakker:2011fu} and {\sc HiggsBounds}~\cite{Bechtle:2011sb, Bechtle:2013wla}
were applied to the posterior samples from the pNMSSM fits to the data on the upper section of Table~\ref{tab.obs}.
The inclusion of {\sc SModelS} constraints during the fits will slow the exploration of the pNMSSM space beyond
tolerance. For this reason, the ``post-processing'' procedure was used. 
The ``post-processing'' means passing the posterior sample points, in SLHA\cite{Skands:2003cj} format, 
to {\sc SModelS} and {\sc HiggsBounds} for imposing the
experimental 95\% confidence limits from 
ATLAS and CMS SUSY bounds~\cite{ATLAS-CONF-2013-024, ATLAS-CONF-2013-047,
  ATLAS-CONF-2013-053, ATLAS-CONF-2013-061, CMS-PAS-SUS-13-018, CMS-PAS-SUS-13-023,
  Chatrchyan:2013lya, Khachatryan:2014qwa, Chatrchyan:2014lfa,
  Khachatryan:2015vra, Chatrchyan:2013xna} and the Tevatron and LHC Higgs physics
bounds~\cite{Bechtle:2011sb, Bechtle:2013wla, arXiv:1402.3051, arXiv:1509.04670,
  arXiv:1402.3244, arXiv:1404.1344, arXiv:1406.7663, arXiv:1504.00936, arXiv:1409.6064, arXiv:1507.05930,
  arXiv:1406.5053, Khachatryan:2014jba, Khachatryan:2014ira, arXiv:1509.00389}
respectively.
The ruled out points were taken out of the samples and the evidence values re-weighted accordingly. 
The impact of {\sc SModelS} and {\sc HiggsBounds} on the evidence values is rather insignificant
since the ruled-out points do not saturate the likelihood space.

\section{Results} 
\subsection{Bayesian evidences}
\begin{table}
\begin{tabular}{|l|l|l|l|l|}
  \hline
  Scenario & $\log_e \mathcal{Z}$ & $K$ & (i/j) Comparison Remarks& Removed constraints \\
  \hline
  ${\cal{H}}_0$ & $-4.99 \pm 0.19$ & $>100$ & (0/1) Strong for ${\cal{H}}_1$   &   \\
  ${\cal{H}}_1$ &$-6.99 \pm 0.22$  & 7.3    & (2/1) Moderate for ${\cal{H}}_2$ &   CDM DD, $Br(B_s \rightarrow \mu^+ \mu^-)$, $Br(B_u \rightarrow \tau \nu)$ \& $\delta a_{\mu}$ \\
  ${\cal{H}}_2$ & $-4.86 \pm 0.18$ & $>100$ & (2/0) Strong for ${\cal{H}}_2$ & \\
  \hline                           
  \hline                           
  ${\cal{H}}_0$ & $-9.99 \pm 0.23$ & 10.0   & (0/1) Moderate for ${\cal{H}}_1$ & \\
  ${\cal{H}}_1$ & $-17.49 \pm 0.19$& $>100$ & (2/1) Strong for ${\cal{H}}_2$   & $Br(B_s \rightarrow \mu^+ \mu^-)$, $Br(B_u \rightarrow \tau \nu)$ \& $\delta a_{\mu}$\\
  ${\cal{H}}_2$ & $-7.24 \pm 0.22$ & $>100$ & (2/0) Strong for ${\cal{H}}_2$ & \\
  \hline                                          
  \hline                                          
  ${\cal{H}}_0$ & $38.24 \pm 0.25$ & 2.7  & (0/1) Inconclusive & \\
  ${\cal{H}}_1$ & $27.64 \pm 0.21$ & 79.0 & (2/1) Moderate for ${\cal{H}}_2$ & None \\
  ${\cal{H}}_2$ & $36.10 \pm 0.23$ & 28.8 & (2/0) Moderate for ${\cal{H}}_2$ & \\
  \hline
\end{tabular}
\caption{Here $\mathcal{Z}$ represents the evidence returned by {\sc PolyChord}. There are three
  set of results demarcated by the double horizontal lines. The first set 
  is for the pNMSSM global fits to the observables shown in Table~\ref{tab.obs} but without
  the CDM direct detection (DD), $Br(B_s \rightarrow \mu^+ \mu^-)$, $Br(B_u \rightarrow \tau \nu)$ and
  $\delta a_{\mu}$ limits. For the second and third sets, the limits from CDM DD searches were added.
  The third set is done with all the observables included. The inclusion of all the limits significantly changed
  the discriminating power of the data.
} 
\label{tab:results}
\end{table}
The results for the Bayesian comparisons between the hypotheses considered are shown in Table~\ref{tab:results}. 
There are three sets of results demarcated by the double horizontal lines.
In the first set, the CDM direct detection (DD), 
$Br(B_s \rightarrow \mu^+ \mu^-)$, $Br(B_u \rightarrow \tau \nu)$ and $\delta a_{\mu}$ limits were not 
included during the fits. For the third set, all the observables were included. 
The Bayesian evidence ($\log_e {\cal Z}$) values returned by {\sc PolyChord} for each of the 
scenarios are shown in the second column.
The Bayes factors, K, for the comparisons between ${\cal H}_i$ and ${\cal H}_j$ scenarios are shown on the third column
while the corresponding remarks are displayed on the fourth column. 
The priors $p({\cal H}_i)$, $i = 0, 1, 2$ were estimated via a random scan of 2156295 pNMSSM points without any of
the ${\cal H}_i$ constraints imposed. Out of these, 1867739, 128, and 7646 points respectively survived 
the ${\cal H}_0$, ${\cal H}_1$, and ${\cal H}_2$ requirements. This way,
$p({\cal H}_0) = 8.6618 \times 10^{-1}$, $p({\cal H}_1) = 5.9361 \times 10^{-5}$, and $p({\cal H}_2) = 3.5459 \times 10^{-3}$. 
As expected, the stronger the data set is, the better the discrimination between 
the scenarios becomes. But this depends on whether all the elements in the set pull the posterior mass towards
  a common region. The addition of two sets of data with tendencies for pulling the posterior in opposite directions
  will dilute the discrimination strength of the combined set. This characteristic behaviour can be seen in going
  from the second to the third set of results shown in Table~\ref{tab:results}. the 
  $Br(B_s \rightarrow \mu^+ \mu^-)$, $Br(B_u \rightarrow \tau \nu)$ and $\delta a_{\mu}$ set tends to prefer lighter SUSY
  states. The inclusion of this set diluted the discrimination power of the combined data set. Without the
$Br(B_s \rightarrow \mu^+ \mu^-)$, $Br(B_u \rightarrow \tau \nu)$ and $\delta a_{\mu}$ limits, the conclusions drawn are
  Moderate or Strong. These changed to Inconclusive or Moderate when the mentioned set of data are included as can be seen
for the third set of results in Table~\ref{tab:results}. For the third set of the results, we discuss the comparisons 
between the scenarios as follows. 
\begin{itemize}
\item The data shows an inconclusive result for the comparison between
  ${\cal H}_0$ against ${\cal H}_1$. This an indication that the CDM DD limits on one hand, versus the 
  $Br(B_s \rightarrow \mu^+ \mu^-)$, $Br(B_u \rightarrow \tau \nu)$ and $\delta a_{\mu}$ set of limits on another are
  sensitive to these scenarios but in opposite directions. 
   This can be seen by noting that the support for ${\cal H}_1$
  against ${\cal H}_0$ changed from Strong to Moderate and then to Inconclusive in going from the
    first to the second and then third set of results shown in Table~\ref{tab:results}.
\item Similarly, the data demonstrates a moderate evidence in support of ${\cal H}_2$ against ${\cal H}_1$.
  For the ${\cal H}_2$ versus ${\cal H}_1$ comparison the evidence changes from Moderate to Strong 
  upon the inclusion of CDM direct detection
  limits. This can be understood as being due to the presence of relatively much lighter $h_1$ in the ${\cal H}_2$ scenario
  (see Fig.~\ref{HiggsMasses}, first-row left plot). Lighter $h_1$ leads to bigger LSP-nucleon
  cross sections and thus more likely to be ruled out by the direct detection limits. 
\item The data has a Moderate support for ${\cal H}_2$ relative to ${\cal H}_0$.
  Over the first two sets of results, the support for ${\cal H}_2$ against ${\cal H}_0$ is Strong. The inclusion of the
    $Br(B_s \rightarrow \mu^+ \mu^-)$, $Br(B_u \rightarrow \tau \nu)$ and $\delta a_{\mu}$ limits diluted the conclusion to a
    Moderate one.
\end{itemize}
All together, there is a Moderate support for ${\cal H}_2$ against  ${\cal H}_0$ and the latter may be
  considered as being ranked first in comparison to ${\cal H}_0$ or ${\cal H}_1$ which can be simultaneously
  ranked second. The hypothesis with a single Higgs boson around $125 \GeV$ is not decisively supported.  The data
has a tendency towards preferring the ${\cal H}_2$ scenario which permits the possibility 
of having an additional but much lighter (than $125 \GeV$) scalar.  

\subsection{Posterior distributions related to CP-even Higgs bosons} 
\begin{figure}[htp]
  \centering 
  \includegraphics[width=.45\textwidth]{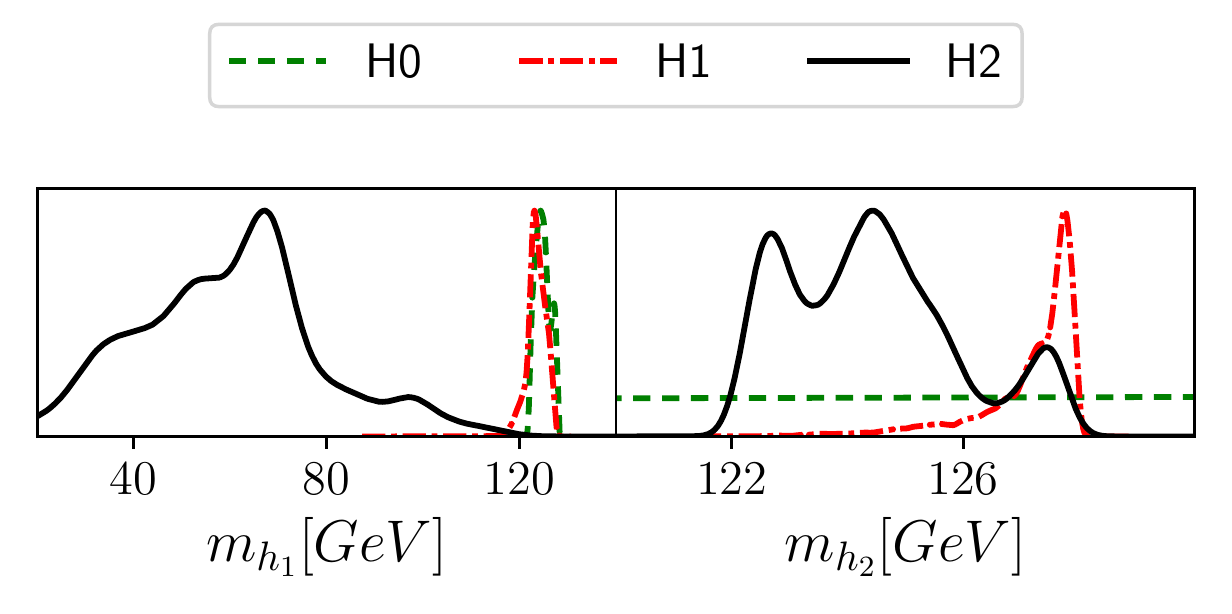}
  \includegraphics[width=.25\textwidth]{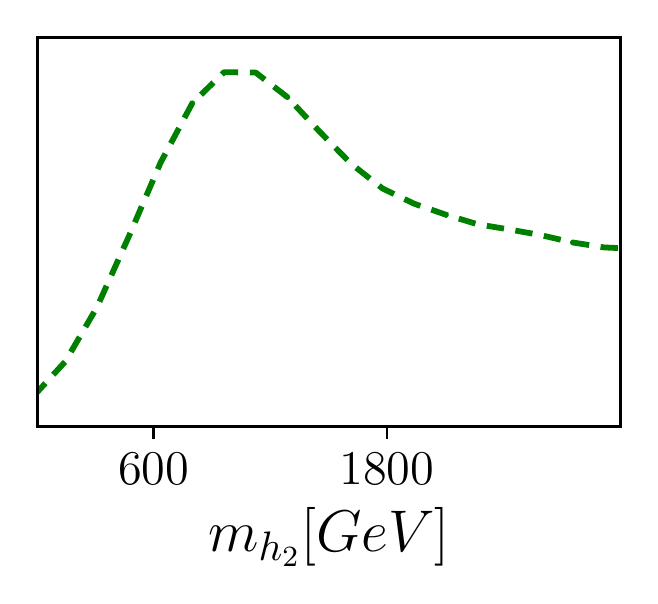}\\
  \includegraphics[width=.45\textwidth]{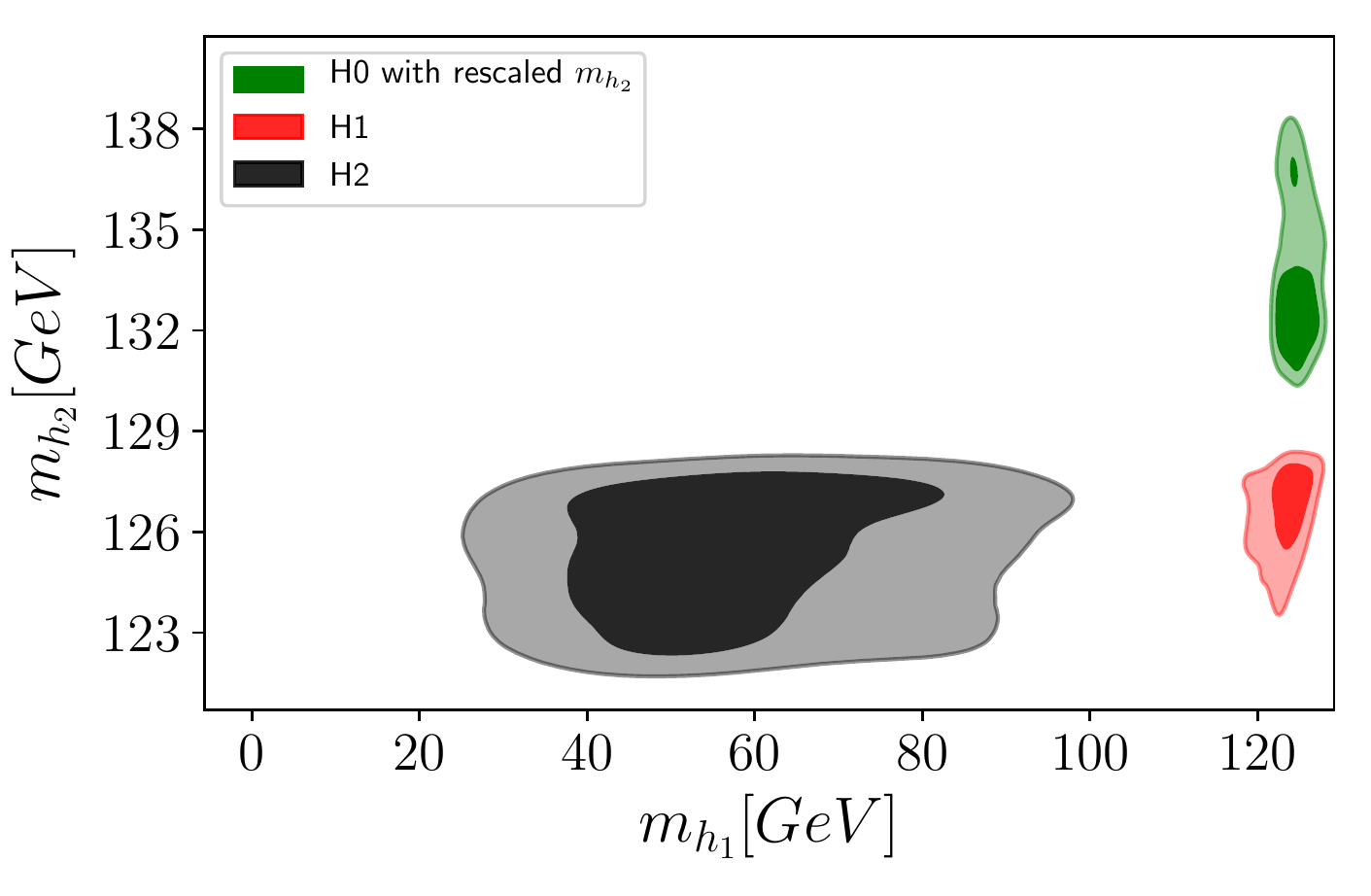}
  \caption{Marginalised one- (top) and two-dimensional (bottom) posterior distributions for the first-two 
    lightest CP-even Higgs bosons. The dashed/green ($H0$),
    dash-dotted/red ($H1$) and solid/black ($H2$) lines represent respectively the 
    hypotheses ${\cal{H}}_0$, ${\cal{H}}_1$ and ${\cal{H}}_2$. The top-right plot shows the $m_{h_2}$ distribution for
    ${\cal{H}}_0$ which is not visible on the top-left plot. 
    For the plot on the bottom, $m_{h_2}$ for
    ${\cal{H}}_0$ scenario ($H0$, on the legend) is re-scaled.
    \label{HiggsMasses}}
\end{figure}
To complement the Bayes' factors reported on Table~\ref{tab:results}, in this subsection we describe the posterior
distributions 
of the first two light Higgs boson masses and couplings. The plots~\footnote{These were are made using GetDist, github.com/cmbant/getdist.}
in Fig.~\ref{HiggsMasses} show respectively the one- and two-dimensional mass distributions 
within each of the scenarios considered. For ${\cal{H}}_2$, $m_{h_1}$
 distribution is peaked at a value much less than 125 GeV. This is because of the interplay between the large electron-positron (LEP) and LHC
constraints. With respect to the position of the peak, the lower mass region is suppressed by LEP constraints such as upper limits on the cross section of 
$e^- \, + \, e^+ \rightarrow h_1 \, Z$ \cite{Buskulic:1993gi, Abdallah:2004wy}. The heavier mass region beyond the peak gets 
ruled out by the upper limits on the reduced production cross sections, at the LHC, such as for
$p\,p \rightarrow h_1 \rightarrow a_1 a_1$~\cite{Khachatryan:2015wka},
$ggF \rightarrow h_1 \rightarrow \gamma \, \gamma$~\cite{ATLAS:2014vga},
and $ggF \rightarrow h_1 \rightarrow \tau \, \tau$~\cite{ATLAS:2014lxa, ATLAS:2016fpj} processes. 

\begin{figure}[htp]
  \centering
  \includegraphics[width=.8\textwidth]{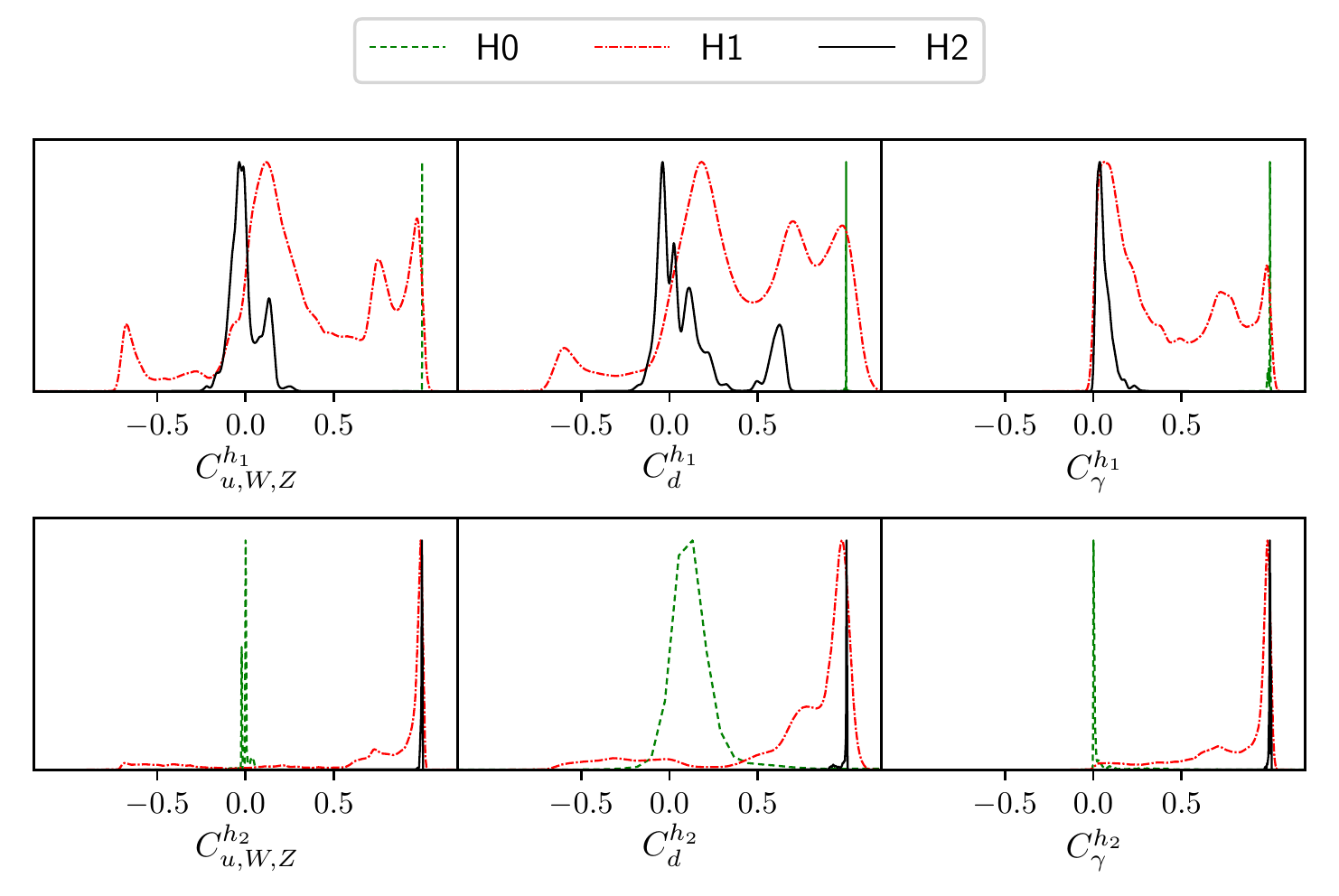}
  \caption{Marginalised posterior distributions for the first-two lightest CP-even Higgs
    bosons ($h_j$, $j = 1$ or $2$) reduced couplings, 
    $C_i^{h_j}$, to matter and gauge particles $i = u, d, V, \gamma$ (respectively up-type, and down-type matter
    particles, W, Z, and photon). The dashed/green line ($H0$), dash-dotted/red line ($H1$) and solid/black
    ($H2$) lines represent respectively the hypotheses ${\cal{H}}_0$, ${\cal{H}}_1$ and ${\cal{H}}_2$.
    \label{hcouplings}}
\end{figure}

The nature of the two lightest CP-even Higgs bosons in each of the three scenarios can be determined by
considering their reduced couplings to fermions (up-type, $u$ and down-type, $d$)
and gauge bosons (W, Z, and $\gamma$). As shown in Fig.~\ref{hcouplings}, 
for ${\cal{H}}_0$ (dashed/green line), $h_1$ is completely SM-like. 
For ${\cal{H}}_1$ (dash-dotted/red line), $h_2$ is mostly SM-like when $h_1$ is not and vice versa. This due to the so-called ``sharing of couplings'' effect.
$h_1$ is almost completely non SM-like within ${\cal{H}}_2$ (solid/black line) for which $h_2$ is
identified as $h_{125}$. These features stem from the combined effects of the various 
limits imposed on the pNMSSM parameter space. The most pronounced of these for $h_1$ 
are the LHC limits on the signal strengths which are directly proportional to the reduced couplings, and subsequently to
the elements of the Higgs mixing matrix. 

\subsection{Posterior distributions related to neutralino CDM candidate}
\begin{figure}[htp]
  \centering
  \includegraphics[width=.8\textwidth]{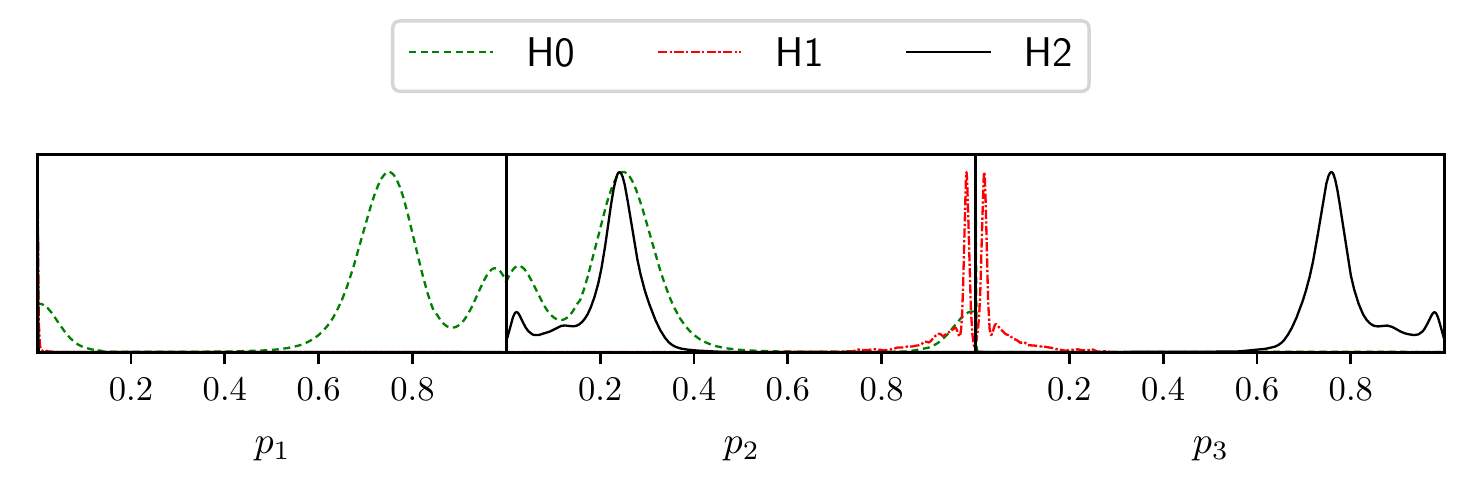}
  \caption{Marginalised posterior distributions for the neutralino LSP content. 
    Here $p_1 = N_{11}^2 + N_{12}^2$, $p_2 = N_{13}^2 + N_{14}^2$,
    and $p_3 = N_{15}^2$ quantifies the gaugino, Higgsino, and singlino content of the LSP.
    The dashed/green ($H0$), dash-dotted/red ($H1$) and solid/black
    ($H2$) lines represent the hypotheses ${\cal{H}}_0$, ${\cal{H}}_1$ and ${\cal{H}}_2$ respectively.
    \label{lspContent}}
\end{figure}

The LSP is identified as a candidate for explaining the
observed CDM relic density. 
Here we show the posterior distributions for the neutralino composition and direct detection cross
sections are presented. The nature of the LSP is one of the most important factors affecting its relic density and 
 scattering cross sections. The relevant part of the Lagrangian for the
neutralino is 
\be
   {\cal L}  =  \frac{1}{2} M_1 \lambda_1 \lambda_1 + \frac{1}{2} M_2 \lambda_2^i \lambda_2^i
   + \frac{1}{2} M_3 \lambda_3^a \lambda_3^a\, .
\ee
Here $\lambda_1$, $\lambda_2^i$ (with $i=1, 2, 3$), and $\lambda_3^a$ (with $a=1, \dots , 8$)
 represents respectively the $U(1)_Y$, $SU(2)_L$ and $SU(3)_c$ gaugino fields. $\lambda_1$ and
$\lambda_2^3$ mix with the neutral Higgsinos $H_1^0, H_2^0, S$ to form a symmetric $5 \times 5$
mass matrix ${\cal M}_0$. With $\psi^0 = (-i\lambda_1 ,
-i\lambda_2^3, H_1^0, H_2^0, S)$, the neutralino mass term takes the form 
${\cal L} = - \frac{1}{2} (\psi^0)^T {\cal M}_0 (\psi^0) + \mathrm{H.c.}$. 
${\cal M}_0$ can be diagonalised by an orthonormal matrix $N_{ij}$ such that the five mass-ordered 
eigenstates are superpositions of $\psi^0_j$:
\be \chi_i^0 = N_{ij} \, \psi^0_j \, .\ee
The neutralino LSP is considered to be gaugino-,
Higgsino- or singlino-like when $p_1 = N_{11}^2 + N_{12}^2$,  $p_2 = N_{13}^2 + N_{14}^2$,
or $p_3 = N_{15}^2$ dominates respectively. The posterior
distributions for these are shown in Fig.~\ref{lspContent}. It resulted in the fact that for ${\cal{H}}_0$
(dashed/green line), the LSP is mixed gaugino-Higgsino with approximately zero singlino content.
The case is different for ${\cal{H}}_2$, for which the LSP is mixed Higgsino-singlino but with dominantly
singlino and zero gaugino content. Instead, in the case for ${\cal{H}}_1$ the LSP is dominantly Higgsino. 

The nature of the neutralino LSP composition determines what leading role the annihilation
and co-annihilation processes (see, e.g., Ref.\cite{Ellwanger:2009dp}) play for
getting the relic density around the experimental value, 
$\Omega_{CDM} h^2 = 0.12$ \cite{Ade:2015xua}. It also determines the processes that could be
involved for the direct detection of the dark matter candidate. Concerning the latter, 
the dominant processes are the t-channel Z or Higgs boson exchange for spin dependent or independent 
interactions, respectively. For instance, highly singlino-like LSP leads to small spin dependent 
cross section. The application of the dark matter direct detection limit will therefore lead to
a posterior distribution with dominantly singlino LSP as is the case within ${\cal{H}}_2$.

\begin{figure}[htp] 
  \centering
  \includegraphics[width=.45\textwidth]{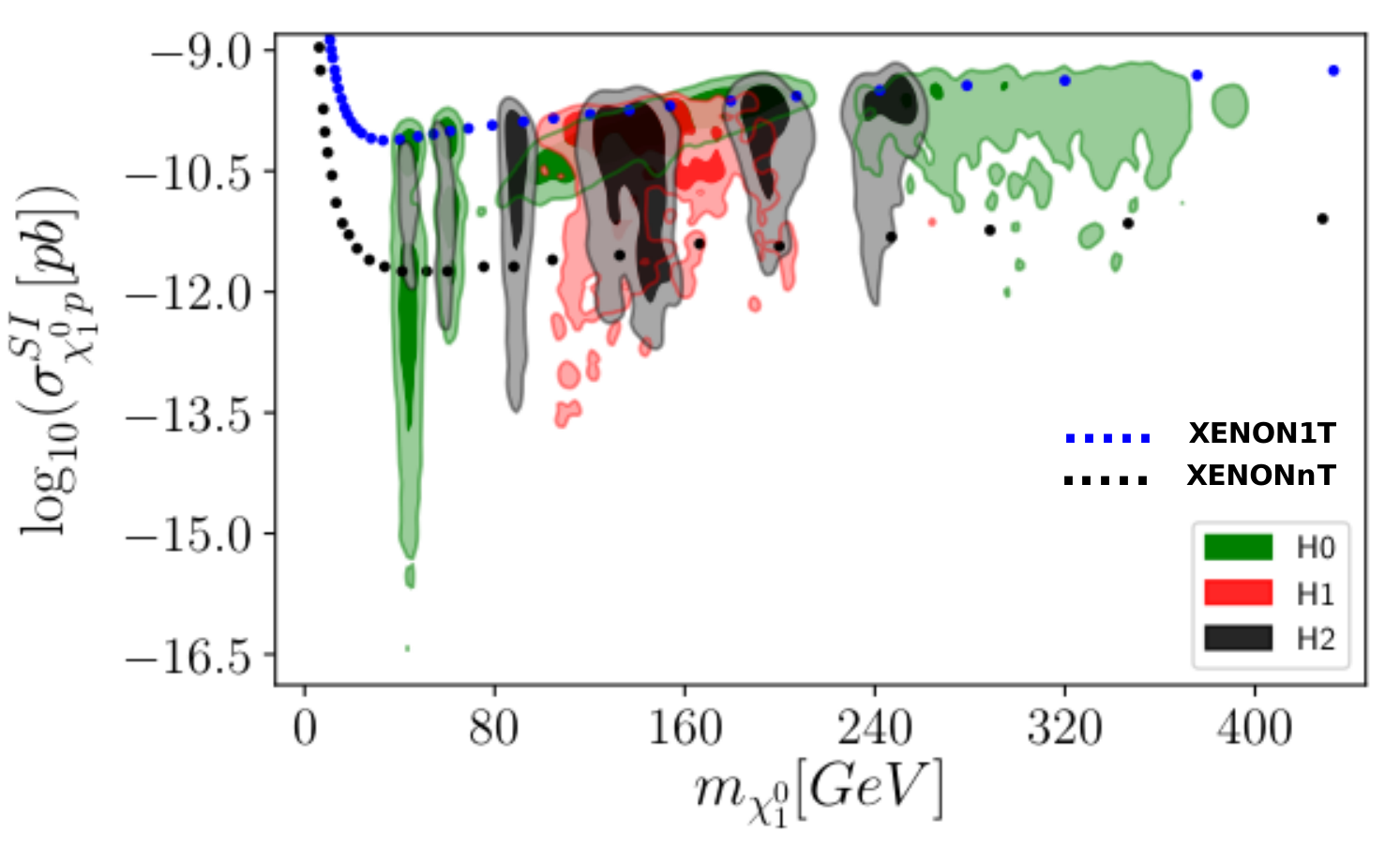}
  \includegraphics[width=.45\textwidth]{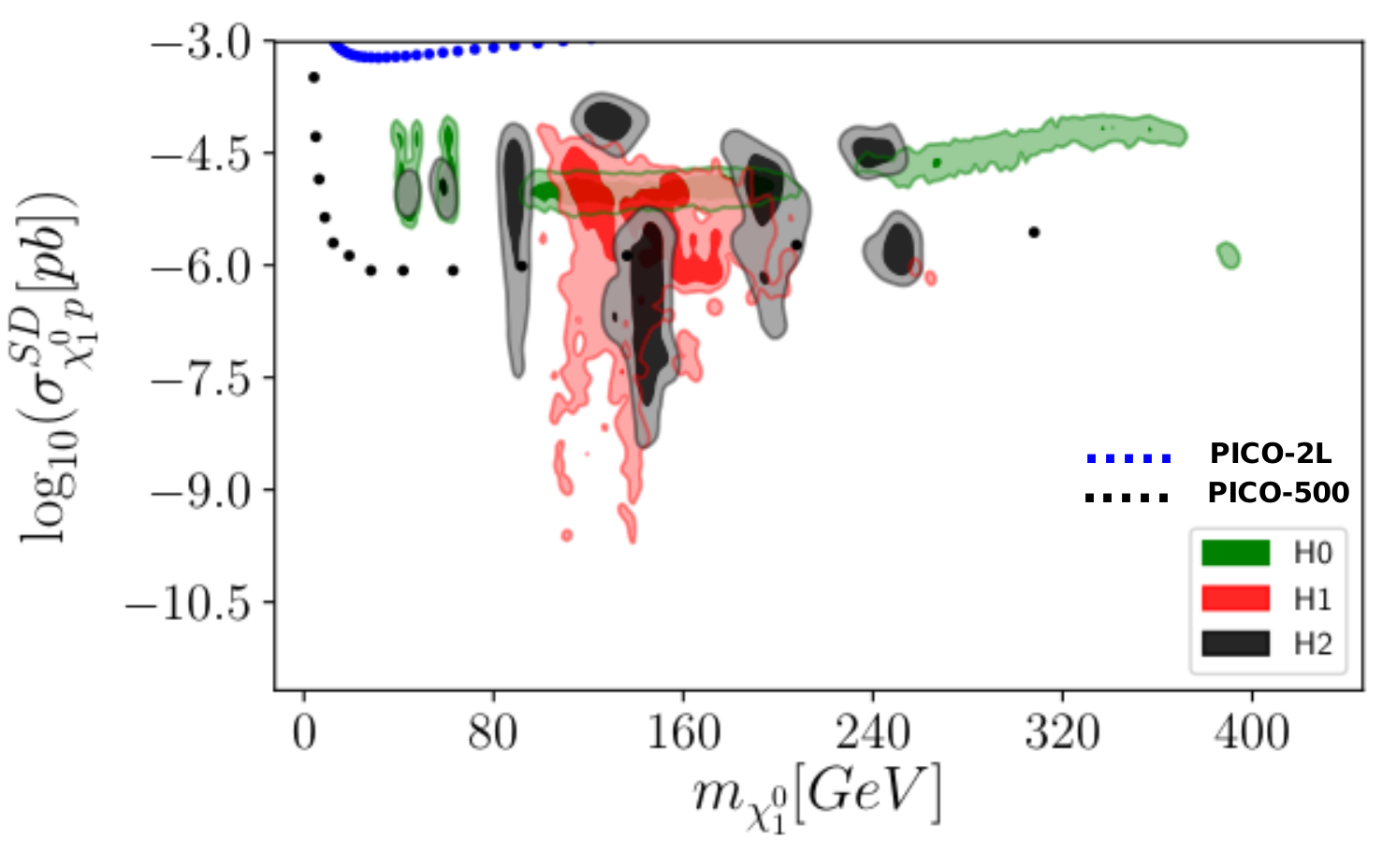}\\
  \includegraphics[width=.35\textwidth]{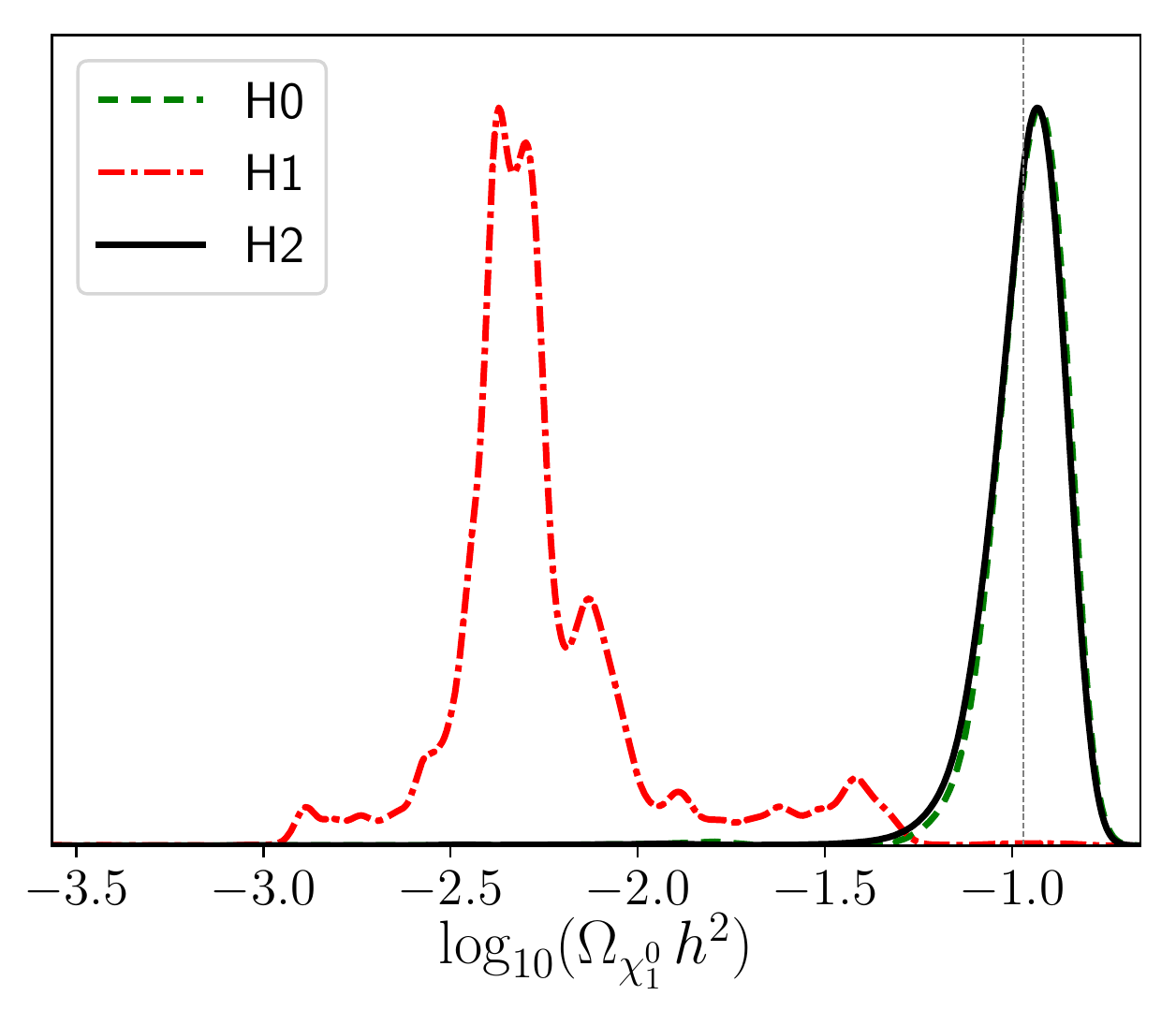}
  \caption{Marginalised two-dimensional posterior distributions for the neutralino LSP 
    mass versus its spin-independent (1st-row left) and spin-dependent (1st-row right)
    scattering cross-section with proton. The green ($H0$), red ($H1$) and black
    ($H2$) regions represent ${\cal{H}}_1$, ${\cal{H}}_1$ and ${\cal{H}}_2$ scenarios respectively. Inner and outer
    contours respectively enclose $68\%$ and $98\%$ Bayesian credibility regions of the
    posteriors. The blue/dotted contour lines on the left- and right-hand side plots show respectively
    the XENON1T~\cite{Aprile:2017iyp} and PICO-2L~\cite{Amole:2016pye} which represents the most constraining 
    of the CDM direct detection limits used.
    The XENONnT and PICO-500 black/dotted contours show possible sensitivity of future upgrades of the
    experiments. The
    posterior distributions of the neutralino relic densities for each of the three pNMSSM scenarios
    are shown on the last plot (3rd-row). The dashed/green, dash-dotted/red and solid/black lines
    respectively represent the ${\cal H}_0$, ${\cal H}_1$ and ${\cal H}_2$ scenarios. For parameter
    points with relic densities (in logarithmic scale) to the left of the mark near $-1.0$, the
    direct detection constraints were rescaled to account for the neutralino dark matter under-production. 
    \label{dmdetection}}
\end{figure}

In Fig.\ref{dmdetection} (first-row) the posterior distributions of the 
spin-independent and spin-dependent LSP-proton scattering cross sections compatible with all the considered collider,
astrophysical, including the CDM direct detection bounds reported
in \cite{Akerib:2016vxi, Aprile:2017iyp, Tan:2016zwf, Amole:2015pla,
  Amole:2016pye, Akerib:2016lao, Fu:2016ega}, and flavour physics constraints are shown.
 The sudden 
suppressions in the direct detection cross sections occur at points for which there
are cancellations from the neutralino-neutralino-Higgs interaction terms~\cite{Ellis:2000ds}. 
The second-row of plots in Fig.\ref{dmdetection} show the DM direct detection limits used for the
global fits. Regions above the contour lines are excluded at 95\% C.L.. Possible sensitivity of the experiments'
future upgrades are also shown (XENONnT and PICO-500). These should probe better the
pNMSSM parameter space, although the situation depends very much on the neutralino CDM relic density.
Under-production of the relic density makes the direct detection limits less constraining.
For instance the regions above the blue/dots line in Fig.\ref{dmdetection} (first-row left) would have
 been excluded. They were not excluded because the relic densities at those points
are much less than the experimentally measured central value around 0.1 as shown in
Fig.\ref{dmdetection} (third/last plot). 

\section{Conclusions}
The nested sampling technique \cite{Skilling} implementation in {\sc PolyChord} \cite{Handley:2015fda, mnrasstv1911}
has been applied for computing the Bayesian evidence within a 26-parameters pNMSSM. 
The evidence values are based on limits from collider, astrophysical bounds on dark matter relic density and direct detection cross
sections, and low-energy observables such as muon anomalous magnetic moment and
flavour physics observables. These were used for comparing between three pNMSSM hypotheses: 
\begin{itemize}
\item ${\cal H}_0$: The scenario for which the observed scalar around $125 \GeV$ is identified as the
  lightest CP-even Higgs boson, $h_1$. $m_{h_1}$ were allowed according to a Gaussian
  distribution with $3 \GeV$ standard deviation.
\item ${\cal H}_1$: This is the same as ${\cal H}_0$ but with the restriction that $m_{h_2}$ be within $122$ to $128 \GeV$. 
\item ${\cal H}_2$: The scenario for which the observed scalar around $125 \GeV$ is identified as the
  second lightest Higgs boson, $h_2$. 
\end{itemize}
Using the Jeffreys' scale for interpreting the evidence values (see Table~\ref{tab:Jeffreys}),
 the analyses indicate that ${\cal H}_2$ could be considered as being 
ranked first, with a Moderate support, amongst the three hypotheses. 
${\cal H}_0$ and ${\cal H}_1$ can be considered as ranked second at the same time. 
That is, the current data used for the Bayesian comparisons favours the hypothesis with
 the possibility of having an additional but much lighter (than $125 \GeV$) scalar. 
The lightest scalar within ${\cal H}_2$ turned out to have mass distribution centred below the W-boson mass as shown
in Fig.~\ref{HiggsMasses}. Due to the ``sharing of couplings'' effect, $h_2$ is SM-like while $h_1$ is dominantly 
singlet-like.

\begin{figure}[htp] 
  \centering
  \includegraphics[width=.6\textwidth]{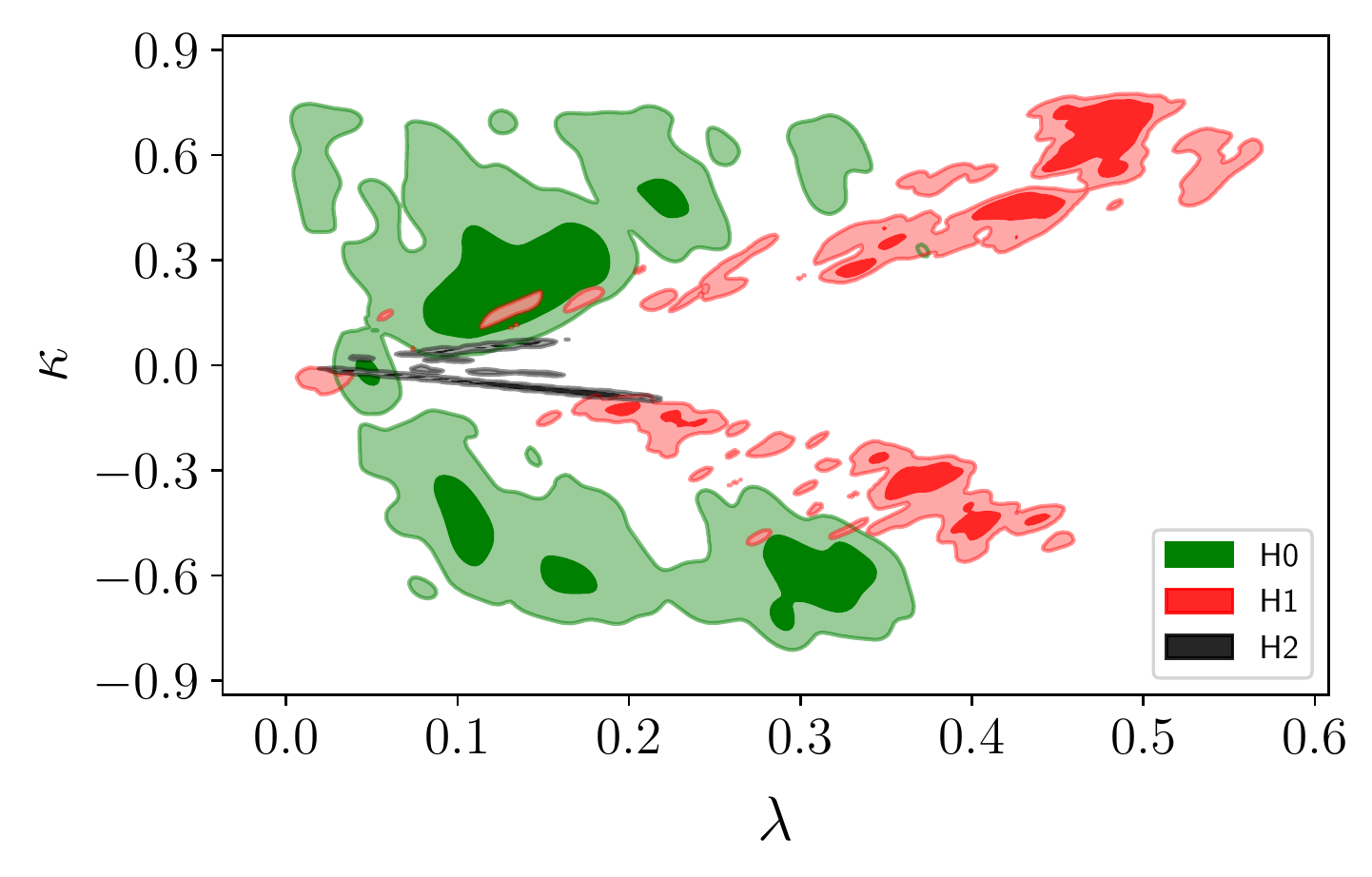}
  \caption{Marginalised posterior distribution for the NMSSM parameters $\lambda$ and $\kappa$.
    The green ($H0$), red ($H1$) and black 
    ($H2$) regions represent the 
    hypotheses ${\cal{H}}_1$, ${\cal{H}}_1$ and ${\cal{H}}_2$, respectively. Inner and outer
    contours respectively enclose $68\%$ and $98\%$ Bayesian credibility regions.
    \label{benchmark}}
\end{figure}
From the posterior distributions presented, pNMSSM benchmark points could be constructed for further analyses 
with regards to non SM-like Higgs bosons. For instance, consider the two-dimensional posterior distribution on the
($\lambda$, $\kappa$) plane shown in Fig.~\ref{benchmark}. For the ${\cal H}_2$ hypothesis, the plane is well
constrained and approximately reduced to a line. The model points along the line could be excellent benchmarks
for testing non-SM Higgs scenarios.

Other possible directions for further investigations can be described as follows.
In this article, the composition of the LSP was not fixed. The only requirement
was that the relic density and the elastic cross section with nucleons be within the experimentally
allowed range. One can go beyond this by demanding, a priori, a particular LSP 
composition. That is, one could require, in addition to what the masses of the light
CP-even Higgs bosons could be, that only a specific LSP composition be allowed during the
pNMSSM parameter space explorations. Next, there are experimental measurements
which could possibly probe, in a better way, the pseudo-degenerate Higgs scenario. These include the
precise determination of the Higgs boson's total decay width. An update of the analyses 
presented here to include these ideas could shed more light about the pseudo-degenerate Higgs
scenario. 

There are caveats within our analyses. One is concerning the 
uncertainty of the Higgs mass prediction. Here we have used the traditional, and possibly too
optimistic uncertainty of $3 \GeV$. A more careful analysis and systematic treatment of the
 uncertainties, see e.g. \cite{Goodsell:2014pla, Staub:2015aea},
 could significantly impact the Bayesian evidence values.
 This is also the case should naturalness priors be used for the analyses.
 In \cite{Athron:2017fxj}, it was shown that imposing naturalness requirements significantly affect 
the evidence. The correlations among the various observables were not included.
Whenever available, the inclusion of correlations could possibly alter the pNMSSM posteriors.
For instance, the measurements of the Higgs boson mass and couplings could come from a single experiment
and therefore likely to be correlated. Finally, the inclusion of SUSY limits, such as in {\sc SmodelS}, 
during the fits and using logarithmic priors should lead to more robust conclusions about the pNMSSM. 

\paragraph*{Acknowledgements:} Thanks to W. Handley for support in
using {\sc PolyChord}; U. Ellwanger, C. Hugonie, and J. Bernon for
{\sc NMSSMTools} related issues; L. Aparicio, M.E. Cabrera and
F. Quevedo for discussions about the NMSSM; and to A. Gatti for
reading and suggestions for improving the manuscript. With support from F. Quevedo, this
work was performed using ICTP's ARGO and the University of Cambridge's Darwin HPC facilities. The 
latter is provided by Dell Inc. using 
Strategic Research Infrastructure Funding from the Higher Education 
Funding  Council for England and funding from the Science and 
Technology  Facilities Council.

\end{document}